\documentclass[preprint,12pt,authoryear]{elsarticle}
\pdfoutput=1
\usepackage{amsmath}
\usepackage{placeins}
\usepackage{pdflscape}
\usepackage{rotating}
\usepackage{color}

\journal{Journal of Theoretical Biology}
\begin{document}

\begin{frontmatter}

\title{The influence of dispersal on a predator-prey system with two habitats}

\author[rvt]{P. Gramlich}
\ead{gramlich@fkp.tu-darmstadt.de}

\author[rvt]{S.J.Plitzko}
\ead{plitzko@fkp.tu-darmstadt.de}

\author[rvt]{L.Rudolf}
\ead{rudolflars@googlemail.com}

\author[rvt]{B.Drossel}
\ead{drossel@fkp.tu-darmstadt.de}

\author[rvt]{T.Gross}
\ead{thilo2gross@gmail.com}

\address{}

\begin{abstract}

\noindent Dispersal between different habitats influences the dynamics and stability of populations considerably. Furthermore, these effects depend on the local interactions of a population with other species. Here, we perform a general and comprehensive study of the simplest possible system that includes dispersal and local interactions, namely a  2-patch 2-species system. We evaluate the impact of dispersal on stability and on the occurrence of bifurcations, including pattern forming bifurcations that lead to spatial heterogeneity, in 19 different classes of models with the help of the generalized modelling approach. We find that dispersal often destabilizes equilibria, but it can stabilize them if it increases population losses. If dispersal is nonrandom, i.e. if emigration or immigration rates depend on population densities, the correlation of stability with dispersal rates is  positive in part of the models. We also find that many systems show all four types of bifurcations and that antisynchronous oscillations occur mostly with nonrandom dispersal.

\end{abstract}


\end{frontmatter}


\section{Introduction}

In ecology both the exploration of dynamical models of food webs \citep{pascual2005ecological,thompson2012food,rooney2012integrating} and the study of spatial metapopulations\citep{holland2008strong,hanski2004ecology,TromeurRudolfGross2013} are well established lines of research. The two modelling approaches emphasize different aspects of ecological dynamics that are both relevant for most species: the dispersal between different habitat patches and tropic interactions with other species. 
Yet, models that combine dispersal with trophic dynamics have only recently begun to appear. In the following we refer to such models that include both these features  as meta-foodwebs. 

An elegant meta-foodweb model that is based on rates for colonization and extinction was proposed by Pillai \citep{pillai2011metacommunity}. More detailed dynamical models were proposed in \citet{Abrams201199}, \citet{Abdllaoui2007335} and \citet{Jansen2001119}, to name a few. The different models draw motivation from different biological systems and hence make different modelling choices regarding the nature of dispersal and local dynamics. For example \citet{Jansen2001119} assumes diffusive dispersal between patches which is appropriate for simple life forms such as bacteria, whereas \citet{Abrams201199} make dispersal dependent on growth rate differences between patches, implying that individuals can actively choose the site with the best growth conditions. Though many of these studies only describe two patches, conclusions from multi-patch models are often consistent with those from two-patch models \citep{JansenDeRoos2000}, and therefore the insights gained from 2-patch systems have wider applications.

 The two mentioned examples for implementing dispersal are only a small subset of the large space of possibilities. In the literature different assumptions are made regarding the functional forms of the number of emigrants from a given habitat patch, the choice of destination, the proportion of survivors  that arrive as immigrants in the destination patch and the settlement success \citep{Amarasekare2008,ArmsworthRoughgarden2008,Rowell2010}. In the simplest case a fixed proportion of the population emigrates per unit time and instantaneously and losslessly settles in a randomly chosen neighbouring patch \citep{Leibold2004}. This type of migration is usually called ``random dispersal'' or ``diffusive migration'', and often leads to a synchronisation of the population dynamics of the two patches \citep{goldwyn2008can,Jansen2001119}. Examples for non-random dispersal are predator evasion and predator pursuit \citep{li2005impact}, or a migration rate that is proportional to the difference in growth rates between two patches \citep{Abrams201199}. 
 
 The type of dispersal strongly affects the stability and the dynamics of the system. In general, more rapid dispersal is more likely to synchronize populations, although 
 synchronisation does not necessarily require strong dispersal \citep{LiebholdKoenigBjornstadt2004}. With adaptive dispersal, antisynchronous oscillations of the two patches are observed, which increase metapopulation persistence. The less similar populations are the less likely is it that dispersal is synchronizing \citep{RantaKaitalaLundberg1998}. Other authors find that increased dispersal can decrease synchrony in population dynamics depending on the interactions between migrating species \citep{KoelleVandermeer2005}. An general investigation of metapopulations based on a linear stability analysis \citep{TromeurRudolfGross2013} found that  costly dispersal and social fencing are stabilizing, while positive density dependence and settlement facilitation reduce stability. Other papers have shown that costly dispersal might be destabilizing to a metapopulation with homogeneous patches \citep{Kisdi2010} so the specific mechanisms appear to be of importance. The effects of dispersal on stability can depend not only on the type but also on the intensity of the dispersal \citep{BriggsHoopes2004}.

While much progress has been made for metapopulations and for specific example systems for meta-foodwebs, a broad and general understanding of how different factors impact the stability of meta-foodwebs is still lacking. For instance is it unclear under which conditions dispersal has a stabilizing impact. Furthermore, meta-foodwebs can potentially undergo various types of instabilities. The study of foodwebs has provided abundant examples of two basic mechanisms of instability. The saddle-node bifurcation, which can lead to the relatively sudden collapse of populations, and the Hopf bifurcation, which gives rise to (at least transient) oscillations. In meta-foodwebs both of these instabilities can occur in two variants. The first of these affects all patches equally and is thus closely related to the bifurcations in non-spatial food web models. In the second type different patches are affected differently. They are thus reminiscent of pattern-forming instabilities such as the Turing and wave-instabilities, which are known from systems of partial differential equations \citep{segel1972dissipative}. While also these bifurcations lead to instability, their impact on the overall population density is less pronounced, and they act as drivers of heterogeneity, which, in the long run, might benefit the system. In addition to these four types of instabilities, meta-foodwebs show further instabilities that occur out of the attractors created by these basic bifurcations, such as bifurcations involving heterogeneous fixed points, and nonlocal bifurcations involving limit cycles or strange attractors.

In order to gain an overview of the possible dynamical patterns of a system and their requirements, the generalized modelling approach \citep{ThiloGross2006,raey,Thilo2009}, which is based on a linear stability analysis of steady states, is particularly powerful. The idea behind this approach is to consider models where the kinetics of some processes have not been restricted to specific functional forms. Considering a model with a specific structure, but containing general functions, allows capturing well-known insights into the structure of the system, without requiring often questionable assumptions on the exact form of kinetics. Further advantages are a  short computation time and ease of biological interpretation. We will confine our study to bifurcations out of homogeneous equilibria. This means that instabilities of heterogeneous systems and nonlocal bifurcations are not considered.

In this paper, we investigate the dynamics of two species on two identical patches using the generalized modelling approach. We are able to analyse a broad class of models that includes several previously studied systems as special cases. We focus on the effect of the type and strength of migration on the stability and the dynamics of the system. We find migration in most cases to be either destabilizing or to  have a marginal effect on stability.
 However, complex migration rules allow for a stabilizing influence of dispersal and can produce saddle-node and Hopf bifurcations and spatial-pattern forming bifurcations.

\section{Model}

\subsection{Generalized modelling formulation}

We consider a  system consisting of two habitat patches,  where each patch $i$ can potentially sustain a prey population $X_i$ and a predator population $Y_i$. We assume a homogeneous system, such that both patches are described by identical parameter values. The population dynamics are described by
\begin{eqnarray}
\dot X_1 & = & G(X_1) - K(X_1)-F(X_1,Y_1)\nonumber\\
         & & + \eta^X E^X(X_2,Y_2,X_1,Y_1)- E^X (X_1,Y_1,X_2,Y_2)\nonumber\\
\dot Y_2 & = & \lambda F(X_1,Y_1)-D(Y_1)\nonumber\\
         & & +\eta^Y E^Y(X_2,Y_2,X_1,Y_1)- E^Y (X_1,Y_1,X_2,Y_2) \label{genmodel}\\
\dot X_2 & = & G(X_2) - K(X_2)-F(X_2,Y_2)\nonumber\\
		& & +\eta^X E^X (X_1,Y_1,X_2,Y_2)- E^X (X_2,Y_2,X_1,Y_1)\nonumber\\
\dot Y_2 & = & \lambda F(X_2,Y_2)-D(Y_2)\nonumber\\
		  & & +\eta^Y E^Y (X_1,Y_1,X_2,Y_2)- E^Y (X_2,Y_2,X_1,Y_1)\, , \nonumber
\end{eqnarray}
 where we used the dot over a variable to indicate the temporal derivative. The variables are in arbitrary units. For the purpose of this paper we will assume that they describe the system in terms of carbon biomass density, however, the same equations also apply to other measures of population, such as abundance. The prey population density changes due to a growth rate $G(X_i)$, a respiration/mortality rate $K(X_i)$, and a rate of biomass loss by predation $F(X_i,Y_i)$. Predator populations have a growth term $\lambda F(X_i,Y_i)$, with the prefactor $\lambda$ describing the efficiency of the energy conversion. The respiration/mortality rate of the predator is given by  $D(Y_i)$. The rate of emigration is $E^U (\textbf{X,Y})$ for both species $U={X,Y}$ and the migration loss factor is $\eta^U$. In the most general case, emigration rates depend on all four populations. The case where the emigration rate of population $U_i$ is proportional to $E_i$ and independent of other variables corresponds to diffusive migration, otherwise we get different versions of adaptive migration.

Modes of the form Eqs.~(\ref{genmodel}) can have multiple feasible steady states, depending one the choice of functional forms and parameter values. In the generalized model we cannot compute the steady states. However, a central insight is that we can still compute conditions for the stability of steady states and express them in the form of meaningful ecological parameters. For this purpose we consider an arbitrary feasible, but not necessarily stable steady state. 

The normalized biomasses are
\begin{equation}
x_i  =  \frac {X_i} {X^{\ast}} \, , \quad {y_i}  =  \frac {Y_i} {Y^{\ast}}\, ,
\end{equation}
and the normalized functions are 
\begin{equation}
 h(\textbf{x,y})  = \frac{H(\textbf{X,Y})}{H(\textbf{X}^{\ast},\textbf{Y}^{\ast})}
\equiv \frac{H(\textbf{X,Y})}{H^{\ast}} 
\end{equation}
with $H=E,F,G,K$ and the asterisk (*) is used to denote the values of variables and functions in this steady state under consideration. 

In terms of these normalized quantities, Eqs.~(\ref{genmodel}) take the form (with $\vec r_1=(x_1,y_1)$ and $\vec r_2 = (x_2,y_2)$)
\begin{eqnarray}
\dot x_1 & = & \frac{G^\ast}{X^\ast}g(x_1) - \frac{K^\ast}{X^\ast}k(x_1)-\frac{F^\ast}{X^\ast}f(x_1,y_1)+\frac{E^{X\ast} \eta^X}{X^\ast}e^X(\vec r_2,\vec r_1)-\frac{E^{X\ast} }{X^\ast} e^X(\vec r_1,\vec r_2)\nonumber\\
\dot y_1 & = & \frac{\lambda F^\ast }{Y^\ast} f(x_1,y_1)-\frac{D^\ast}{Y^\ast}d(y_1)+\frac{E^{Y\ast} \eta^Y}{Y^\ast}e^Y(\vec r_2,\vec r_1)-\frac{E^{Y\ast} }{Y^\ast} e^Y(\vec r_1,\vec r_2)\label{gennorm}\\
\dot x_2 & = & \frac{G^\ast}{X^\ast}g(x_2) - \frac{K^\ast}{X^\ast}k(x_2)-\frac{F^\ast}{X^\ast}f(x_2,y_2)+\frac{E^{X\ast} \eta^X}{X^\ast}e^X(\vec r_1,\vec r_2)-\frac{E^{X\ast} }{X^\ast} e^X(\vec r_2,\vec r_1)\nonumber\\
\dot y_2 & = & \frac{\lambda F^\ast }{Y^\ast} f(x_2,y_2)-\frac{D^\ast}{Y^\ast}d(y_2)+\frac{E^{Y\ast} \eta^Y}{Y^\ast}e^Y(\vec r_1,\vec r_2)-\frac{E^{Y\ast} }{Y^\ast} e^Y(\vec r_2,\vec r_1).\nonumber
\end{eqnarray}
It is now useful to identify the total biomass turnover rate of the populations at their steady state. At the steady state, all the functions in (\ref{gennorm}) take the value 1, and the gain and loss terms are equal and can be denoted as
\begin{eqnarray}
\alpha^X & = & \frac{G^{\ast}}{X^{\ast}} +\frac{E^{X{\ast}} \eta^X}{X^{\ast}}= \frac{K^{\ast}}{X^{\ast}}+\frac{F^{\ast}}{X^{\ast}}+\frac{E^{X{\ast}} }{X^{\ast}}\, , \\
\alpha^Y & = & \frac{\lambda F^{\ast} }{Y^{\ast}} +\frac{E^{Y{\ast}} \eta^Y}{Y^{\ast}}=\frac{D^{\ast}}{Y^{\ast}}+\frac{E^{Y{\ast}} }{Y^{\ast}}\,.
\end{eqnarray}
If the variables describe abundances then the parameter $\alpha^X$ is the total turnover rate for the prey in a patch. For instance,  a value of 0.25/year would indicate that an individual spends on average 4 years in the patch before either dying or emigrating. 

It is convenient to measure time in terms of the inverse of the prey turnover rate $\alpha^X$. In these rescaled  time units the turnover rate of the prey is 1 and the turnover rate of the predator is
\begin{equation}
\alpha =  \frac{\alpha^Y}{\alpha^X}\, .
\end{equation}
Furthermore, we denote the different relative contributions to the growth and loss terms by so-called \textit{scale parameters}  $\delta$, $\nu$ and $\rho$ and their complements
\begin{eqnarray}
 \delta & = &\frac{1}{\alpha^X \tilde{\rho}^X}\frac{F^{\ast}}{X^{\ast}} \, , \qquad \tilde\delta  =  (1-\delta)=\frac{1}{\alpha^X \tilde{\rho}^X} \frac{K^{\ast}}{X^{\ast}}\, , \\
 \nu^U & = &\frac{1}{\alpha^U} \frac{E^{U{\ast}} \eta^U}{U^{\ast}} \, , \qquad \tilde\nu^U  =  (1-\nu^U)=\frac{1}{\alpha^U}\frac{\{G^{\ast};\lambda F^{\ast} \}}{U^{\ast}}\, , \\
 \rho^U & = & \frac{1}{\alpha^U} \frac{E^{U{\ast}}}{U^{\ast}} \, , \qquad \tilde\rho^U=(1-\rho^U)=\frac{1}{\alpha^U}\frac{\{K^{\ast}+F^{\ast};D^{\ast}\}}{U^{\ast}}\, .
\end{eqnarray} 

Scale parameters  describe the branching of the biomass flow, i.e., they are proportions of the total biomass influx or output attributed to a specific function or mechanism. The parameter $\delta$ denotes the proportion of energy intake of the prey that is eventually lost again due to respiration or mortality, whereas the $\tilde{\delta}$ is the proportion of the energy intake that is eventually lost due to predation. The parameters $\nu$ and $\tilde\nu$ denote the relative contributions of migration and  feeding to the total gain, respectively. The parameters $\rho$ and $\tilde\rho$ are the counterpart to $\nu$ and $\tilde\nu$ and denote the relative loss by emigration and by the within-patch processes (respiration/mortality and predation). A value of zero of a scale parameter means no biomass flow attributed to a mechanism (e.g. $\delta$=0 means no gain by predation) and value of 1 means a biomass flow completely dominated by a mechanism (e.g. $\tilde{\rho}^X$=1  means loss only by predation and respiration).

Using the turnover rates and the scale parameters, we can write Eq.~(\ref{gennorm}) as
\begin{eqnarray}
\dot x_1 & = & \alpha^X[\tilde\nu^X g(x_1) - \tilde\rho^X_1 \tilde\delta k(x_1) -  \tilde\rho^X \delta f(x_1,y_1) + \nu^X e^X(\textbf{x,y}) - \rho^X e^X(\textbf{x,y})]\nonumber\\
\dot y_1 & = & \alpha^Y[\tilde\nu^Y f(x_1,y_1) - \tilde\rho^Y d(y_1) + \nu^Y e^Y(\textbf{x,y}) - \rho^Y e^Y(\textbf{x,y})] \label{genend}\\
\dot x_2 & = & \alpha^X[\tilde\nu^X g(x_2) - \tilde\rho^X\tilde\delta k(x_2) -  \tilde\rho^X \delta f(x_2,y_2) + \nu^X e^X(\textbf{x,y}) - \rho^X e^X(\textbf{x,y})]\nonumber\\
\dot y_2 & = & \alpha^Y[\tilde\nu^Y f(x_2,y_2) - \tilde\rho^Y d(y_2) + \nu^Y e^Y(\textbf{x,y}) - \rho^Y e^Y(\textbf{x,y})].\nonumber
\end{eqnarray}
For analysing the stability we linearise the system around the steady state under consideration. In the normalized system the steady state is at x=y=1. The linearisation can then be expressed in terms of the Jacobian matrix. For a system of four dynamical variables this is a $4\times4$ matrix defined by 
\begin{equation}
J_{i,j} = \left. \frac{\partial  \dot{V_i}}{\partial V_j}\right|_*,
\end{equation} 
where $V=(x_1,y_1,x_2,y_2)$ is the set of state variables, and $|_*$ indicates that the derivatives are evaluated in the steady state (1,1,1,1).

For the case of a system with two identical patches the $4 \times 4$ matrix can be written as a block matrix 

\begin{center}
\begin{align}
J=\left(
   \begin{array}{cc}
     L & M  \\
     M & L  
   \end{array}\right)\, 
   \end{align}
   \end{center}
with a matrix $L$ that describes the in-patch dynamics,
\begin{align}
L=\left(
   \begin{array}{cc}
     1 & 0  \\
     0 & \alpha  
   \end{array}\right)\left(
   \begin{array}{cc}
     \tilde\nu^X \phi - \tilde\rho^X \tilde\delta \mu^X - \tilde\rho^X \delta \gamma + \nu^X \hat\omega^X - \rho^X \omega^X  & - \tilde\rho^X \delta \psi + \nu^X \hat\kappa^X - \rho^X \kappa^X  \\
     \tilde\nu^Y \gamma + \nu^Y \hat\kappa^Y - \rho^Y \kappa^Y  & \tilde\nu^Y \psi - \tilde\rho^Y \mu^Y + \nu^Y \hat\omega^Y - \rho^Y \omega^Y 
   \end{array}
\right)
\end{align}
and a matrix $M$ that captures between-patch dynamics,
\begin{align}
M=\left(
   \begin{array}{cc}
     1 & 0  \\
     0 & \alpha  
   \end{array}\right)\left(
   \begin{array}{cc}
      \nu^X \omega^X - \rho^X \hat\omega^X  & \nu^X \kappa^X - \rho^X \hat\kappa^X  \\
      \nu^Y \kappa^Y - \rho^Y \hat\kappa^Y  & \nu^Y \omega^Y - \rho^Y \hat\omega^Y  
   \end{array}
\right) \, .
\end{align}
The new parameters in these matrices  are called \emph{exponent parameters}, and they are the derivatives of the population dynamic functions with respect to the prey or predator populations. The definitions of the exponent parameters are
\begin{align}
 \phi:&=\frac{\partial g(x)}{\partial x}\Bigr|_* \, ,\qquad \mu^X:=\frac{\partial k(x)}{\partial x}\Bigr|_* \, ,\qquad \mu^Y:=\frac{\partial d(y)}{\partial y}\Bigr|_*  \, ,\nonumber \\ 
  \gamma:&=\frac{\partial f(x,y)}{\partial x}\Bigr|_* \, ,\qquad \psi:=\frac{\partial f(x,y)}{\partial y}\Bigr|_* \, , \nonumber\\ 
 \hat\omega^X:&=\frac{\partial e^X(x_1,y_1,x_2,y_2)}{\partial x_2}\Bigr|_* \, ,\qquad \omega^X:=\frac{\partial e^X(x_1,y_1,x_2,y_2)}{\partial x_1}\Bigr|_* \, , \label{eq:ExponentDef}\\ 
 \hat\kappa^X:&=\frac{\partial e^X(x_1,y_1,x_2,y_2)}{\partial y_2}\Bigr|_* \, ,\qquad \kappa^X:=\frac{\partial e^X(x_1,y_1,x_2,y_2)}{\partial y_1}\Bigr|_* \, ,\nonumber\\
 \hat\omega^Y:&=\frac{\partial e^Y(x_1,y_1,x_2,y_2)}{\partial y_2}\Bigr|_* \, ,\qquad \omega^Y:=\frac{\partial e^Y(x_1,y_1,x_2,y_2)}{\partial y_1}\Bigr|_* \, ,\nonumber\\
 \hat\kappa^Y:&=\frac{\partial e^Y(x_1,y_1,x_2,y_2)}{\partial x_2}\Bigr|_* \, ,\qquad \kappa^Y:=\frac{\partial e^Y(x_1,y_1,x_2,y_2)}{\partial x_1}\Bigr|_* \nonumber\, .
\end{align}
The exponent parameters are so-called elasticities, i.e.~they are logarithmic derivatives of the original functions. For example
\begin{equation}
\phi = \left. \frac{\partial g(x)}{\partial x} \right|_* = \left. \frac{\partial{\rm log}(G(X)) }{\partial {\rm log}(X)}\right|_* . 
\end{equation} 
Such logarithmic derivatives have a number of advantageous properties. Elasticities as a measure of nonlinearity were first introduced in the 1920s in economic theory, because of their statistical properties which allows them to be estimated precisely based on limited noisy data. For the same reason these parameters are now commonly used in metabolic control theory. 

In generalized models we use elasticities mainly because they allow  for an intuitive interpretation: For any linear function, say $G(X)=AX$, the corresponding elasticity $\partial {\rm log} (G(X))/\partial {\rm log}(X) |_1 $ is 1, regardless of the slope $A$. More generally, for any power law $G/(X) = AX^p$ the elasticity is the power law exponent $p$. For more complex functions the elasticity provides an intuitive measure of the degree of saturation of the underlying process. For example for the Holling type-II kinetics the corresponding elasticity is close to 1 in the linear regime at low prey density, but approaches zero at high prey density, where predators are saturated.
 \begin{table}[htb]
\begin{center}
 \begin{tabular}{c| p{4cm}|c|p{5.5cm}}
Parameter& Interpretation & Range & Examples\\
\hline
 $\phi$& Elasticity of the gain function & $[0,1]$ & 0: growth independent of population size, e.g. due to nutrient limitation, 1: growth prop. to population density\\
$\mu^X$& Sensitivity of prey mortality to prey population & $[1,2]$ & 1: constant mortality 2: mortality proportional to density, e.g. due to diseases \\
$\mu^Y$& Sensitivity of predator mortality to predator population & $[1,2]$& same as for prey\\
 $\gamma$ & Sensitivity of predation gain to prey population & $[0,2]$& Holling type functions: value close to 1 for small prey population, low value for large pop. due to saturation\\
 $\psi$ & Sensitivity of predation gain to predator population & $[0,1]$& 1: No predator competition, lower values are due to predator interference \\
 \end{tabular} 
 \caption{The five exponent parameters that characterize in-patch dynamics, their meaning, and their typical range.}
  \label{tab:ExponentExp}
 \end{center}

 \end{table}

The $\omega$ and $\kappa$ are the migration exponent parameters, where the exponent $\omega$ describes the dependence of the emigration rate on the density of the emigrating population, whereas $\kappa$ captures  the dependence of emigration on the density of the other species in the patch, respectively. The parameters with a hat $\hat{\omega},\hat{\kappa }$ are defined analogously but describe the dependence on the densities in the destination patch. Different types of adaptive migration imply different parameter ranges for these exponent parameters. Simple diffusive migration for species $U$ means $\omega^U=1$, and when the emigration rate increases with population size we obtain an exponent $\omega^U>1$. In the case of predator evasion, prey emigration increases with predator density, which means a value $\kappa^X>0$. In predator pursuit, predator emigration decreases with prey density, which means  $\kappa^Y<0$.

The remaining five parameters (first five in Eq.~(\ref{eq:ExponentDef})) describe the elasticity of the local (within-patch) processes. These parameters have been discussed extensively in previous work (e.g. \citet{ThiloGross2006}). Hence, we summarize their interpretation in Tab.~\ref{tab:ExponentExp}.
\FloatBarrier

\subsection{Linear stability, eigenvalues, and bifurcations}
\label{sec:MSF}

In order to evaluate the stability of a steady state,
we have to calculate the eigenvalues of the Jacobian at  the steady state. 
Due to the symmetric block structure of the Jacobian, the eigenvalue equation can be solved with the ansatz\citep{macarthur2008symmetry}
\begin{displaymath}
\left(
   \begin{array}{cccc}
     L_1 & L_2 & M_1 & M_2  \\
     L_3 & L_4 & M_3 & M_3  \\
     M_1 & M_2 & L_1 & L_2   \\
     M_3 & M_4 & L_3 & L_4  
   \end{array}\right)\left(
   \begin{array}{c}
     \xi_1  \\ 
     \xi_2  \\ 
     \pm \xi_1  \\ 
     \pm \xi_2  
   \end{array}
\right)=\lambda\left(
   \begin{array}{c}
     \xi_1  \\ 
     \xi_2  \\ 
     \pm\xi_1  \\ 
     \pm\xi_2  
   \end{array}
\right)\, ,
\end{displaymath}
with $L_{i}$ and $M_{i}$ denoting the matrix elements of $L$ and $M$.
This is equivalent to solving the 2x2 problem
\begin{displaymath}
\left(
   \begin{array}{cc}
     L_1\pm M_1 & L_2\pm M_2   \\
     L_3\pm M_3 & L_4\pm M_4    
   \end{array}\right)\left(
   \begin{array}{c}
     \xi^{\pm}_1  \\ 
     \xi^{\pm}_2  
   \end{array}
\right)=\lambda\left(
   \begin{array}{c}
     \xi^{\pm}_1  \\ 
     \xi^{\pm}_2  
   \end{array}
\right)\, .
\end{displaymath}
 Every solution of the eigenvalue equation describes an eigenmode of the system, i.e., a specific perturbation that retains its shape while it grows or declines in time. For the plus sign, the eigenvector components relating to the two patches are identical, for the minus sign, they point in opposite directions.

 The steady state
  is stable if the real parts of all eigenvalues are negative. The eigenvalues for a 2x2 matrix $J'$ can be written in terms of  trace (T) and the determinant ($\Delta$) of the matrix as
\begin{equation}
 \lambda_{1,2}=\frac{1}{2}T(J')\pm\sqrt{\frac{1}{4}T(J')^2-\Delta (J')}\, .
 \label{eq:eigenvalue}
\end{equation}
 In order to obtain a stable steady state, the real part of both eigenvalues must be negative, which is the case when the trace is negative and the determinant is positive. These criteria must be satisfied for both matrices $L\pm M$ (below denoted by the index "+" for "L+M" and "-" for "L-M").\\
  As parameters are changed, a stable steady state can become unstable when the parameter change causes an eigenvalue to cross the imaginary axis and acquire a positive real part. For our system, there are four different  types of such local bifurcations.
    
Let us first consider the $"J_+=L+M"$ case. Instabilities that are detected by the analysis of this matrix affect both patches equally and in synchrony. First, stability of the steady state can be lost in a saddle-node bifurcation, which occurs when $\Delta (J_+)=0$ and $T(J_+)<0$ while both eigenvalues of $J_-$ have a negative real part. 
In this bifurcation a sudden change in the population densities occurs, which most likely results in the collapse of one or both populations in both patches simultaneously. The second fundamental bifurcation in which stability can be lost is the Hopf bifurcation. This bifurcation occurs when $T(J_+)=0$ and $\Delta (J_+)>0$ while the eigenvalues of $J_-$ have a negative real part. The Hopf bifurcations detected in the L+M matrix gives rise to synchronous oscillations. From the study on non-spatial predator-prey systems it is well known that oscillation amplitudes can grow rapidly after the bifurcation, leading to subsequent extinctions \citep{RosenzweigMcArthur1963}. 

Let us now consider the $J_-=L-M$ matrix. Bifurcations detected in this matrix affect the two patches in opposite directions. Again, stability can be lost either in a saddle-node or in a Hopf bifurcation. However, now the
Saddle-node bifurcation leads to a shift where each population increases in one patch and decreases in the other. This occurs when $\Delta (J_-)=0$ and $T(J_-)<0$ and both eigenvalues of $J_+$ have a negative real part.  
A Hopf bifurcation detected in the $L-M$ matrix leads to anti-synchronous oscillations. While this can lead to large-amplitude oscillations in both individual patches, the overall biomass in the system will stay nearly constant as the loss in one patch is compensated by gains in the other. This type of bifurcation occurs at $T(J_-)=0$ and $\Delta (J_-)>0$ with both eigenvalues of $J_+$ having a negative real part.

Both bifurcations occurring in the $L-M$ matrix can be considered as \emph{pattern-forming} bifurcations that create spatial heterogeneity. In fact, the saddle-node bifurcation in the $L-M$ matrix is closely reminiscent of the Turing bifurcation in systems of partial differential equations, which gives rise to stationary patterns. The Hopf bifurcation in the $L-M$ matrix is reminiscent of the wave instability bifurcation in partial differential equations, which leads to travelling waves. 

In the following we will use the terms Hopf and saddle-node bifurcation only for the corresponding bifurcations from the $L+M$ matrix. For simplicity we will denote the respective bifurcations in the $L-M$ system as Turing and wave instability. We emphasize that this is a loose usage of the bifurcation names that we adopt here as it leads to the right ecological intuition although it is not strictly mathematically justified\footnote{Strictly, a bifurcation is of a given type only if it can be reduced to the type's normal form. For instance the saddle-node bifurcation observed in the $L-M$ system can be reduced to the saddle-node normal form but not to the Turing normal form. So strictly it is a saddle-node and not a Turing bifurcation. Still one can justify denoting this bifurcation as 'Turing' as follows: The bifurcation would not change fundamentally when we considered a system with more than two patches. The bifurcation would then be governed by a matrix that is closely reminiscent of the network laplacian, which in turn can be seen as a discretization of the real space laplacian on a complex network. In this sense even the 2-patch system can be interpreted as a discretization of an underlying continuous space in which the bifurcation would be a true Turing bifurcation.}.

\subsection{Classes of models studied in the following}
In order to evaluate the proportion of stable systems and the frequency of the different types of bifurcations, we need to specify intervals for the exponent parameters and the scale parameters. Different classes of models are characterized by different choices for these intervals. We use several models from the existing literature as well as more general models. All these models are listed in Tab.~\ref{tab:BigTabular}.

\FloatBarrier

\pagebreak

 \begin{sidewaystable}[htb]
\begin{center}
\tiny{
 \begin{tabular}{cccccccccccccccccccc}\hline
 Scen. & $\phi$ & $\delta$ & $\gamma$ & $\psi$  & $\mu^X$ & $\mu^Y$ & $\alpha$ & $\nu^X$ & $\nu^Y$ & $\rho^X$ & $\rho^Y$ & $\omega^X$ & $\omega^Y$ & $\hat\omega^X$ & $\hat\omega^Y$ & $\kappa^X$ & $\kappa^Y$ & $\hat\kappa^X$ & $\hat\kappa^Y$\\\hline\hline
 (1) Stand.& $[0;1]$ & $[0;1]$ & $[0;2]$ & $[0;1]$ & $[1;2]$ & $[1;2]$ & $[10^{-3};10^{1}]$  & $[0;1]$  & $[0;1]$  & $[0;1]$  & $[0;1]$  & $[-2;2]$  & $[-2;2]$  & $[-2;2]$  & $[-2;2]$  & $[-2;2]$  & $[-2;2]$  & $[-2;2]$  & $[-2;2]$  \\
 (1x) Stand.& $[0;1]$ & $[0;1]$ & $[0;2]$ & $[0;1]$ & $[1;2]$ & $[1;2]$ & $[10^{-3};10^{1}]$  & $[0;1]$  & $[0;1]$  & $0$  & $0$  & $[-2;2]$  & $[-2;2]$  & $[-2;2]$  & $[-2;2]$  & $[-2;2]$  & $[-2;2]$  & $[-2;2]$  & $[-2;2]$  \\
 (1z) Stand.& $[0;1]$ & $[0;1]$ & $[0;2]$ & $[0;1]$ & $[1;2]$ & $[1;2]$ & $[10^{-3};10^{1}]$  & $[0;1]$  & $[0;1]$  & $\nu^X$  & $\nu^Y$  & $[-2;2]$  & $[-2;2]$  & $[-2;2]$  & $[-2;2]$  & $[-2;2]$  & $[-2;2]$  & $[-2;2]$  & $[-2;2]$  \\\hline
 (2) Diff & $[0;1]$ & $[0;1]$ & $[0;2]$ & $[0;1]$ & $[1;2]$ & $[1;2]$ & $[10^{-3};10^{1}]$  & $[0;1]$  & $[0;1]$  & $[0;1]$  & $[0;1]$  & $1$  & $1$  & $0$  & $0$  & $0$  & $0$  & $0$  & $0$  \\
 (2x) Diff & $[0;1]$ & $[0;1]$ & $[0;2]$ & $[0;1]$ & $[1;2]$ & $[1;2]$ & $[10^{-3};10^{1}]$  & $[0;1]$  & $[0;1]$  & $0$  & $0$  & $1$  & $1$  & $0$  & $0$  & $0$  & $0$  & $0$  & $0$  \\
 (2z) Diff & $[0;1]$ & $[0;1]$ & $[0;2]$ & $[0;1]$ & $[1;2]$ & $[1;2]$ & $[10^{-3};10^{1}]$  & $[0;1]$  & $[0;1]$  & $\nu^X$  & $\nu^Y$  & $1$  & $1$  & $0$  & $0$  & $0$  & $0$  & $0$  & $0$  \\
 (2y) Diff & $[0;1]$ & $[0;1]$ & $[0;2]$ & $[0;1]$ & $[1;2]$ & $[1;2]$ & $[10^{-3};10^{1}]$  & $[0;1]$  & $0$  & $\nu^X$  & $0$  & $1$  & $1$  & $0$  & $0$  & $0$  & $0$  & $0$  & $0$  \\\hline
 (3) Mchich& $1$ & $1$ & $1$ & $1$ & $0$ & $1$ & $[10^{-3};10^{1}]$  & $[0;1]$  & $[0;1]$  & $[0;1]$  & $[0;1]$  & $1$  & $1$  & $0$  & $0$  & $[0;1]$  & $0$  & $0$  & $0$  \\
 (3x) Mchich& $1$ & $1$ & $1$ & $1$ & $0$ & $1$ & $[10^{-3};10^{1}]$  & $[0;1]$  & $[0;1]$  & $0$  & $0$  & $1$  & $1$  & $0$  & $0$  & $[0;1]$  & $0$  & $0$  & $0$  \\\hline
 (4) Jans95 & $1$ & $[0;1]$ & $[0;1]$ & $1$ & $2$ & $1$ & $[10^{-3};10^{1}]$  & $[0;1]$  & $[0;1]$  & $[0;1]$  & $[0;1]$  & $1$  & $1$  & $0$  & $0$  & $0$  & $0$  & $0$  & $0$  \\
 (4x) Jans95 & $1$ & $[0;1]$ & $[0;1]$ & $1$ & $2$ & $1$ & $[10^{-3};10^{1}]$  & $[0;1]$  & $[0;1]$  & $0$  & $0$  & $1$  & $1$  & $0$  & $0$  & $0$  & $0$  & $0$  & $0$  \\
 (4z) Jans95 & $1$ & $[0;1]$ & $[0;1]$ & $1$ & $2$ & $1$ & $[10^{-3};10^{1}]$  & $[0;1]$  & $[0;1]$  & $\nu^X$  & $\nu^Y$  & $1$  & $1$  & $0$  & $0$  & $0$  & $0$  & $0$  & $0$  \\\hline
 (5) Jans01 & $1$ & $[0;1]$ & $[0;1]$ & $1$ & $2$ & $1$ & $[10^{-3};10^{1}]$  & $0$  & $[0;1]$  & $0$  & $[0;1]$  & $0$  & $1$  & $0$  & $0$  & $0$  & $0$  & $0$  & $0$  \\
 (5z) Jans01 & $1$ & $[0;1]$ & $[0;1]$ & $1$ & $2$ & $1$ & $[10^{-3};10^{1}]$  & $0$  & $[0;1]$  & $0$  & $\nu^Y$  & $0$  & $1$  & $0$  & $0$  & $0$  & $0$  & $0$  & $0$  \\
 (6) Huang & $1$ & $[0;1]$ & $[0;1]$ & $1$ & $2$ & $1$ & $[10^{-3};10^{1}]$  & $0$  & $[0;1]$  & $0$  & $[0;1]$  & $0$  & $1$  & $0$  & $0$  & $0$  & $[-1;0]$  & $0$  & $0$  \\
 (6R) Huang & $1$ & $[0;1]$ & $[0;1]$ & $1$ & $2$ & $1$ & $[10^{-3};10^{1}]$  & $0$  & $[0;1]$  & $0$  & $[0;1]$  & $0$  & $1$  & $0$  & $0$  & $0$  & $[0;1]$  & $0$  & $0$  \\
 (7) Abrams11 & $1$ & $[0;1]$ & $[0;1]$ & $1$ & $2$ & $1$ & $[10^{-3};10^{1}]$  & $0$  & $[0;1]$  & $0$  & $[0;1]$  & $0$  & $1$  & $0$  & $0$  & $0$  & $-\hat\kappa^Y$  & $0$  & $[0;\frac{\lambda b}{4 h}]$  \\\hline
 (8) El Abdl. & $1$ & $[0;1]$ & $1$ & $1$ & $2$ & $1$ & $[10^{-3};10^{1}]$  & $[0;1]$  & $[0;1]$  & $[0;1]$  & $[0;1]$  & $1$  & $1$  & $0$  & $0$  & $1$  & $-1$  & $0$  & $0$  \\\hline
 (9) Growth & $[0;1]$ & $[0;1]$ & $[0;2]$ & $[0;1]$ & $[1;2]$ & $[1;2]$ & $[10^{-3};10^{1}]$  & $[0;1]$  & $[0;1]$  & $[0;1]$  & $[0;1]$  & $\phi$  & $\psi$  & $\phi$  & $\psi$  & $\gamma$  & $0$  & $\gamma$  & $0$  \\
 (10) Growth2 & $1$ & $[0;1]$ & $[0;1]$ & $1$ & $2$ & $1$ & $[10^{-3};10^{1}]$  & $[0;1]$  & $[0;1]$  & $[0;1]$  & $[0;1]$  & $\phi$  & $\psi$  & $\phi$  & $\psi$  & $\gamma$  & $0$  & $\gamma$  & $0$  \\\hline
\end{tabular}
}
\caption{The different models (``scenarios'') analysed in this paper, together with the intervals from which the parameters were chosen randomly. The models \emph{Jan95}, \emph{Jan01}, \emph{Abrams11}, \emph{Mchich} \emph{ElAbdl.} can be found in \citet{Jansen95}, \citet{Abdllaoui2007335}, \citet{Abrams201199}, \citet{Jansen2001119} and \citet{Mchich2007343}.\emph{Stand} represents the most general scenario where all parameters take the biologically plausible range. \emph{Diff} combines general local dynamics with diffusive migration. \emph{Growth} couples growth rates and feeding terms to migration terms. \emph{Jans95}, \emph{Jans01}, \emph{Huang} and \emph{Abrams11} share the same local dynamics but have different migration mechanisms. \emph{Mchich} and \emph{ElAbdll.} do not share local dynamics with any other scenario and have to treated separately. In the text and the figures, the labels in the left column (number plus letter, or name) are used to name the scenarios. The letter $x$, $y$ and $z$ after the number indicates a  migration rule where there is no migration loss ($x$), or where gain and loss are identical, $\nu_{x} = \rho_{x}$, which means that all migrating individuals arrive at their destination ($y$ if this applies only to the prey and $z$ if this applies to both species). }
 \label{tab:BigTabular}

\end{center}
\end{sidewaystable}

\pagebreak

\FloatBarrier 

Group (1) represents the most general models, for which all exponent parameters have their maximum meaningful range. In group (2), migration is  diffusive for both species, but the other local exponent parameters are still the in maximum meaningful range. Models (4) to (7) have a Holling type 2 functional response with logistic growth and no predator interference, and they differ with respect to the migration term. Model (4) \emph{Jans95} shares the local patch dynamics with most of the models in the literature \citep{RosenzweigMcArthur1963} and has a simple diffusive migration. Model (5) is the same as model (4) but with no prey migration. Model (6R) \emph{Huang} adds that over-abundance of prey facilitates predator migration while model (6) incites predators to stay if there is plenty of food available. Model (7) \emph{Abrams11} implements predator pursuit, which means that predators migrate towards the patch with larger prey abundance. 

Models (9) and (10) 
establish a relation between consumption rate and migration: the migration rate of the predator scales in the same way as its feeding rate (i.e., the biomass production), and the migration rate of the prey scales with its own growth rate as well as with the predator feeding rate and with predation on the other patch.

We group all these models into two classes, with class I comprising the models with general intervals for the five in-patch scale parameters  (models (1), (2), (9)) and class II comprising models (4) to (7) and (10), which are based on the Rosenzweig-MacArthur model. 

Models (3) \emph{Mchich} and (8) \emph{ElAbdllaoui} fix more parameters for the in-patch dynamics but have interesting migration rules: 
\emph{Mchich}  lets the prey flee if there are many predators in the own patch. \emph{ElAbdllaoui} additionally makes the predator sedentary if there is plenty of food available.

\section{Results}

\subsection{Proportion of stable systems}

We evaluated the proportion of stable states in an ensemble where the scale and exponent parameters were chosen at random from the intervals indicated in Tab.~\ref{tab:BigTabular}. We generated $10^7$ random sets of parameters for each model where each parameter was drawn uniformly from the respective intervals. We define the proportion of stable webs (PSW) as the number of parameter sets which produce a stable steady state, divided by the total number of sets sampled. We use a random distribution of values for exponent and scale parameters as we do not assume any relationship between parameters. A specific scenario will have most likely dependencies  between functions (e.g. predator feeding rate and prey mortality share common variables) but these are subsets of our parameter space. Weights for the parameter distribution would skew the results towards particular assumptions which we want to avoid.
In the absence of migration (i.e., for single patches), we obtained PSW=0.939 for class I, PSW=0.639 for class II and PSW=0 for  model \emph{Mchich}
  and PSW=1 for model \emph{ElAbdllaoui}. These results are a measure of the stability of the within-patch dynamics alone. We know from \citet{levins1974discussion,Thilo2009} that mortality and feeding terms tend to be positively correlated with stability and growth terms are negatively correlated with stability for local dynamics. This means that high values for $\gamma$ and $\mu^{x,y}$ and low values for $\phi$ and $\psi$ enhance stability.  
 Based on this knowledge we expect  model \emph{Mchich} to be less stable than  model \emph{ElAdbllaoui}, and systems with class I local dynamics to be more stable than systems with class II local dynamics. This is confirmed by our results.  Model \emph{Mchich} is a special case as the particular parameter choices of \citet{Mchich2007343} create a pair of purely imaginary eigenvalues, both for the case with and without migration. This does not allow for a conclusive linear stability analysis, and higher order terms are needed to judge the stability of a steady state.

  The PSW values obtained in the presence of migration are given in Tab.~\ref{tab:AllCorr}.   In all cases, stability with migration is smaller than or roughly equal to stability without migration. The proportion of stable webs ranges from 0.05 for model (10), which has a complex migration rule, to 0.94 for class I with diffusive, conservative migration, i.e., models (2y) and (2z). In fact, the models (1), (9), (10) have the largest drop in stability due to migration. These models have in common that $\hat \omega$ and/or $\hat \kappa$ are nonzero, which means that dispersal of a population depends on the population of the other species on the other patch. This can be understood by applying the qualitative stability considerations established by \citet{levins1974discussion}: Including in the Jacobian an element that connects different species on different patches creates a positive feedback loop of length 3, which has a strong destabilizing effect. 
        
It is interesting to note that the class II systems are less stable than the models with diffusive migration, (2) to (2z), despite the high exponent of closure $\mu_{x} = 2$. A high exponent of closure is well known to be stabilizing since it implies a strong density limitation of the predator\citep{levins1974discussion,Thilo2009,Plitzko2012}. In order to investigate the effect of the exponent of closure, we run the class II systems also with  $\mu_{x} = 1$ instead of 2. In this case all systems became unstable. When $\mu_{x}$ was chosen at random from the interval $[1,2]$, the average stability was lower than for the isolated patches (around 0.50), but with trends similar to those for $\mu_{x} = 2$. 
 
\begin{table}[htb]
\begin{center}
 \begin{tabular}{c|c|c}
 Scenario & PSW                &               PSW     \\
          & without migration  &          with migration\\
\hline
\hline 
 (1)Stand  & 0.939& 0.227\\
  (1x)Stand & 0.939& 0.570\\
   (1z)Stand & 0.939& 0.394\\
    \cline{1-3}
  (2)Diff  & 0.939& 0.924\\
  (2x)Diff  & 0.939& 0.876\\
   (2y)Diff  & 0.939& 0.942\\
    (2z)Diff  & 0.939& 0.939 \\
   	\cline{1-3}     
   	 (9)Growth & 0.939& 0.450 \\
      \hline
      \hline
     (4)Jans95 & 0.693& 0.693\\
     (4x)Jans95  & 0.693& 0.693 \\ 
     (4z)Jans95  & 0.693&  0.693\\
   \cline{1-3}
     (5)Jans01 & 0.693 & 0.693 \\
     (5z)Jans01  & 0.693& 0.693 \\
   \cline{1-3}
     (6)Huang  & 0.693& 0.552 \\
     (6R)Huang &  0.693& 0.530\\
     (7)Abrams11  & 0.693& 0.693\\    
     (10)Growth2  & 0.693& 0.051\\
    \hline
    \hline        
     (3)Mchich & 0.000& 0.00 \\
     (3x)Mchich  & 0.000& 0.00 \\
     \hline
     \hline
     (8)ElAbdllaoui  & 1.000 & 0.625\\

\end{tabular}
 \end{center}
 \caption{The different scenarios and the proportion of stable webs with and without migration. The double horizontal lines separate the classes I, II, and the exceptions model \emph{Mchich} and model \emph{ElAdbllaoui}. The parameter values for the different scenarios were drawn uniformly from the intervals given in Tab.~\ref{tab:BigTabular}.}
 
 \label{tab:AllCorr}
 \end{table}

\begin{figure}[htb]
\centering
 \includegraphics[width=\textwidth]{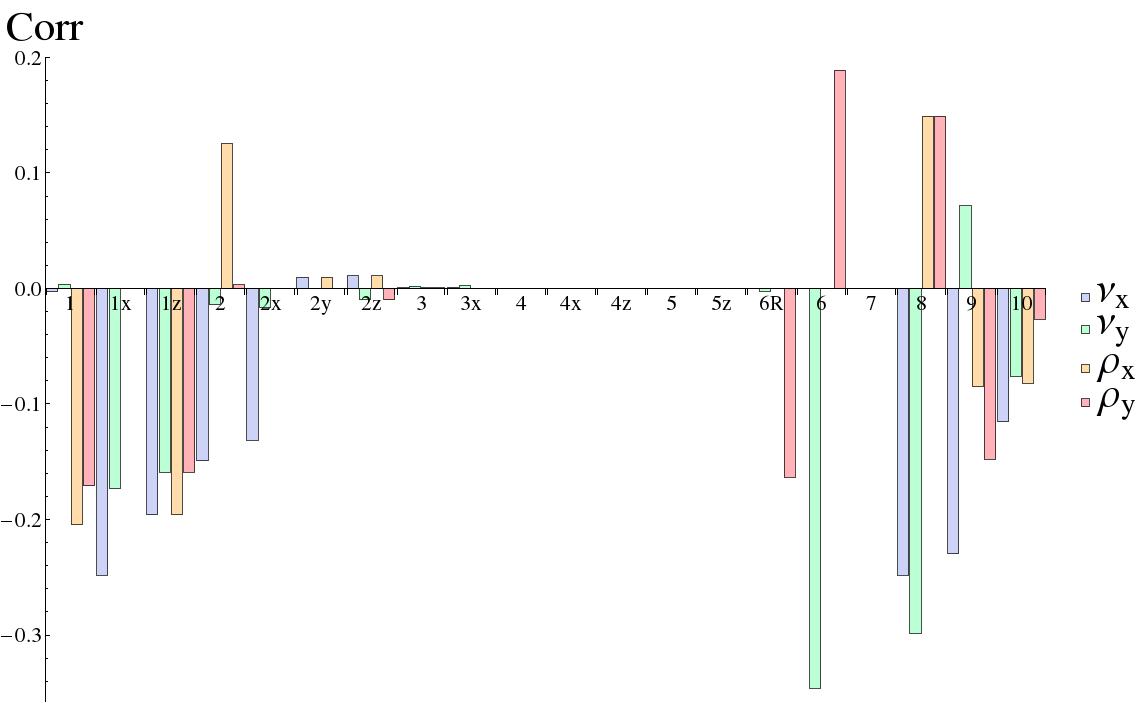} 
  \caption{The correlation of the four scale parameters of migration ($\nu_{x},\nu_{y},\rho_{x},\rho_{y}$) with stability for $10^{7}$ randomly chosen parameter sets for each model. The different models are arranged along the x axis, the correlation value is given along the y axis. The different scale parameters are coded by colour. Most models show a negative correlation of the migration parameters with stability.}
  \label{fig:AllModCorr}
\end{figure}

In order to obtain more detailed information about the effect of migration on stability, we evaluated the correlation of stability with the scale parameters of migration. We used the Pearson product-moment correlation coefficient to compute the correlation between stability and the migration scale parameters. 
 The result is shown in Fig.~\ref{fig:AllModCorr}.

\FloatBarrier 
 
In most cases, the correlation is zero or negative, and there are only four instances of a significant positive correlation with migration. Positive correlation values mean that higher relative migration rates make the steady state
 more stable, negative correlation values indicate that the steady state
  becomes less stable. It is interesting to note that out of the five instances with positive correlations four are loss terms $\rho_{x,y}$. This fits together with the observation that loss terms are comparable to death terms, which are known to be stabilizing for local dynamics \citep{Thilo2009}.

To summarize, the results in  Fig.~\ref{fig:AllModCorr} imply that most models become less stable when the contribution of migration to the total gain and loss terms increases. However, several models become more stable with increasing migration losses, \emph{ElAbdllaoui}, \emph{Huang} and \emph{Diff}, and only  model \emph{Growth} becomes more stable with increasing migration gains. The models \emph{ElAbdllaoui},\emph{Huang} and \emph{Growth} all show types of adaptive migration. The model \emph{Diff} has simple diffusion as a dispersal mechanism, though both \emph{Diff} and \emph{Growth} have wide parameter interval ranges for local parameters.

\FloatBarrier

\begin{figure}[htb]
\centering
 \includegraphics[width=\textwidth]{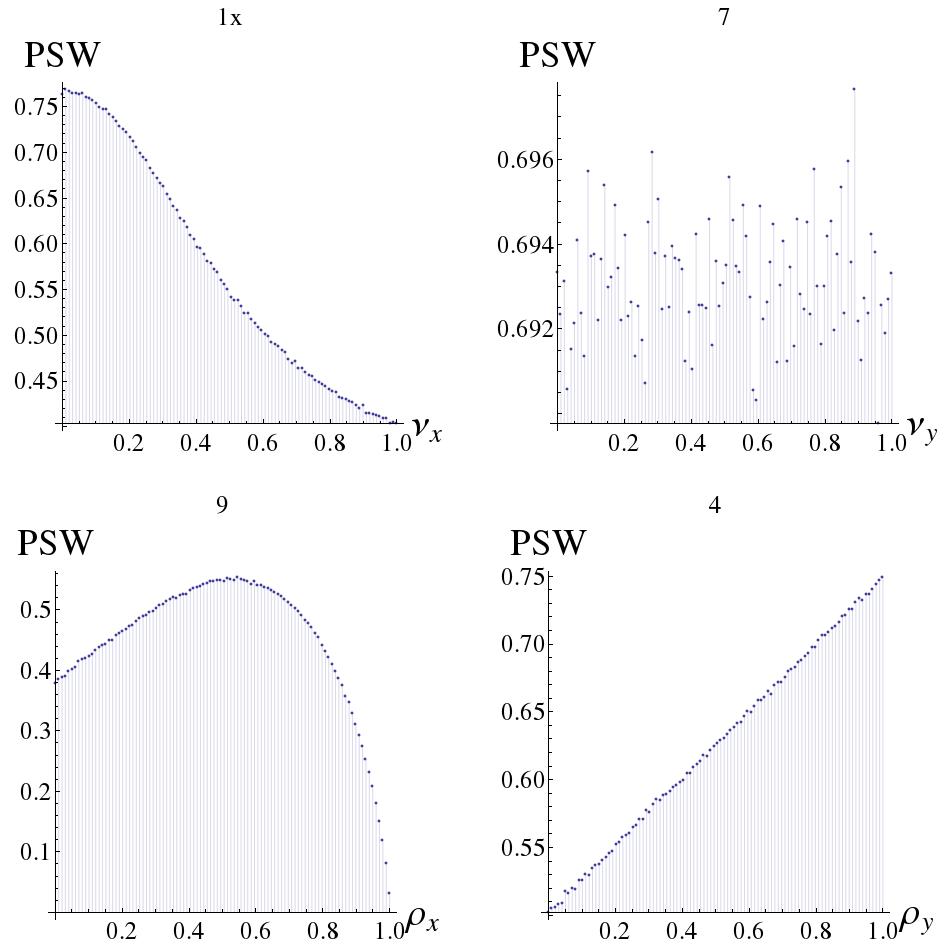} 
 \caption{
 Examples of the different types of functional relation between scale parameters and average stability. The graphs show the proportion of stable systems as function of a migration scale parameter. The numbers above each plot denote the model (see Tab.~\ref{tab:BigTabular}) which is defined by the intervals of the exponent parameters and the relationship between scale parameters used to create the data. Apart from a monotonous decrease or increase, there are also models with a constant proportion of stable systems and those with maxima at intermediate values of the scale parameter.  }
 \label{fig:SampleHistogram}
\end{figure}

In addition to evaluating the correlation of the scale parameters with stability, we also evaluated how average stability changes as a scale parameter is varied. Example curves that represent the qualitatively different types of behaviour found are shown  in Fig. \ref{fig:SampleHistogram}.  

 The top left graph shows a  monotonous decrease, which is most often obtained and applies to most scenarios that have a negative correlation of the scale parameter with stability.  The bottom right graph shows a monotonous increase, which is seen in the cases of positive correlation of the scale parameters with stability.  The bottom left graph shows an instance of a non-monotonous curve, found for the scenarios \emph{Stand}, \emph{Growth} and \emph{Growth2}, which have complex migration rules, where more than two of the eight migration exponent parameters ($\omega$ , $\kappa$ ) are different from zero. This shows that cross-patch cross-species interactions as seen in these three classes can create positive correlation of migration scale parameters with stability within certain ranges even if the overall correlation remains negative. 
The top right graph is an instance where  migration has no effect on stability.

\FloatBarrier

Inspired by these findings, we investigated additional cases: Instead of linear relationships for the sensitivity of the migration on the respective population ($\omega_{x,y}$=1, as used in class II), we used quadratic relationships ($\omega_{x,y}$=2) for the scenarios of class II. This caused clear trends, with the average stability decreasing with increasing values of the scale parameters $\nu_{x,y}$ and increasing with $\rho_{x,y}$. Even though slope and absolute values vary, this effect dominates over every other influence on stability and shapes the curves almost solely. This is understood by realizing that $\omega_{x,y}=2$ implies large loss terms at high values of $\rho$. The dependence on $\nu$ is more complicated as $\nu$ affects only the non-diagonal elements of the Jacobian matrix. High $\nu$ creates larger diagonal elements for the migration sub-matrix. This increases the trace $T(J_{+})$ and thus decreases the likelihood of a stable steady state even though $T(J_{-})$ becomes more stable as the total stability is limited by the stability of each sub-matrix.

\subsection{Example of analytical computation}

\begin{figure}[htb]
\centering
 \includegraphics[width=70mm]{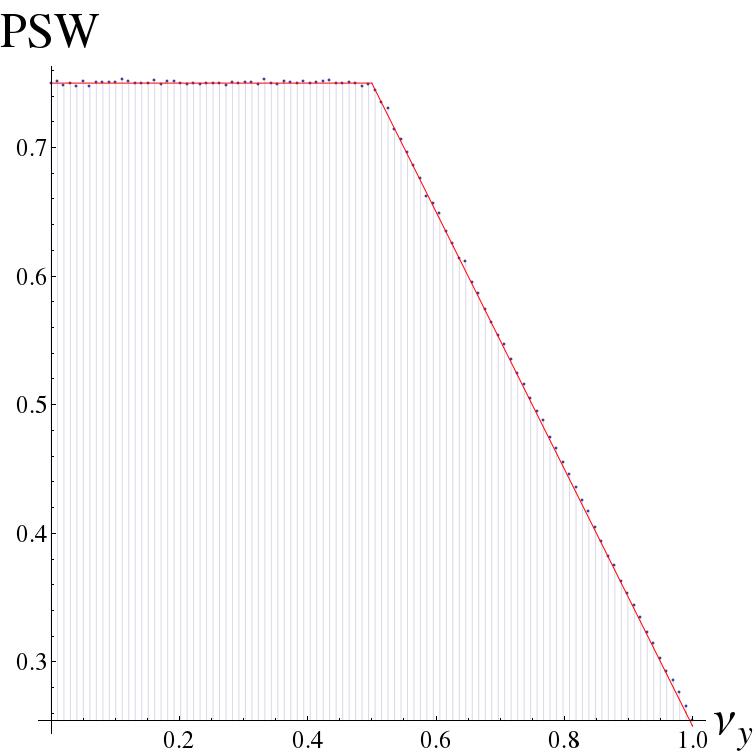} 
 \caption{The average stability as a function of  $\nu_{y}$ for the scenario \emph{ElAbdllaoui}. The solid red line shows the analytical result.}
 \label{fig:ExampleElAbd}
\end{figure}

In most cases, the stability curves are not easily understood. The stability condition comprises four inequalities, two for each matrix $L \pm M$ (see paragraph after eq. (\ref{eq:eigenvalue})). Each of these inequalities contains several parameters, and checking whether they are satisfied requires the consideration of a multitude of cases. In the following we demonstrate for the model with the smallest number of free parameters, which is the model \emph{ElAbdllaoui}, how the shape of the stability curve results from the inequalities. As an example we use the average stability versus the scale parameter $\nu_{y}$, as it has a distinct shape of two linear sections with different slopes which can be seen in Fig. \ref{fig:ExampleElAbd}.

 The traces and determinants for this model are

\begin{eqnarray*}
T(J_{+}) & = & 1-2(1-\delta )(1-\rho_{x} )-\delta (1-\rho_{x} )-\rho_{x} \, ,\\
T(J_{-}) & = & 1-2(1-\delta )(1-\rho_{x} )-\delta (1-\rho_{x} )-\rho_{x} - 2 \nu_{x} - 2\alpha_{y}\nu_{y}\, ,\\ 
\Delta (J_{+}) & = & -(\delta (-1+\rho_{x} )-\rho_{x} +\nu_{x} )(\alpha_{y}(1+\rho_{y} -\nu_{y}) -\alpha_{y}\nu_{y})\, ,\\
\Delta (J_{-}) & = & -2 \alpha_{y} (1-2(1-\delta )(1-\rho_{x} )-\delta (1-\rho_{x} )-\rho_{x} -2\nu_{x} )\nu_{y} \, ,\\
			&  & -(\delta (-1+\rho_{x} )-\rho_{x} -\nu_{x} )(\alpha_{y} (1+\rho_{y} -\nu_{y} )+\alpha\nu_{y} ) \, .
\end{eqnarray*}  
The steady state is stable if both traces are negative and both determinants are positive. The values of $\delta $, $\rho_{x,y} $ and $\nu_{x,y} $ are in the interval $[0,1]$. This means that $T(J_{+,-}) <0 $ and $\Delta(J_{-}) >0 $. The stability is thus solely  dependent on $\Delta (J_{-})$.   Rewriting $\Delta (J_{+})$ as
$$
\Delta (J_{+})  =   -\alpha_y f_{1}\cdot f_{2}$$
with $f_{1}  =  \delta(\rho_{x} -1)-\rho_{x}+\nu_{x}$ and $f_{2}  =  1+\rho_{y}-2\nu_{y}$, the stability condition becomes $f_1f_2< 0$, which means that $f_1$ and $f_2$ must have opposite signs. The term $f_1$  is independent of $\nu_{y}$ and thus cannot change sign as $\nu_{y}$ is changed. With all parameters in $f_1$ being random numbers in $[0,1]$,  $f_{1}<0$ in 75\% of the cases, and  $f_{1}>0$ in 25\% of the cases. $f_{2}$ is always larger than zero as long as $\nu_{y}<0.5$. This means that the steady state is stable in 75\% of the cases. As $\nu_{y}$ increases from $0.5$ to 1, the probability that  $f_{2}>0 $ drops linearly from 1 to 0. For  $\nu_{y}=1$, only 25\% percent of the systems are stable because now $f_1$  must be positive. This explains the linear drop of the stability curve from 0.75 to 0.25.

\FloatBarrier

\subsection{Bifurcations}

\begin{figure}[htb]
\centering
 \includegraphics[width=\textwidth]{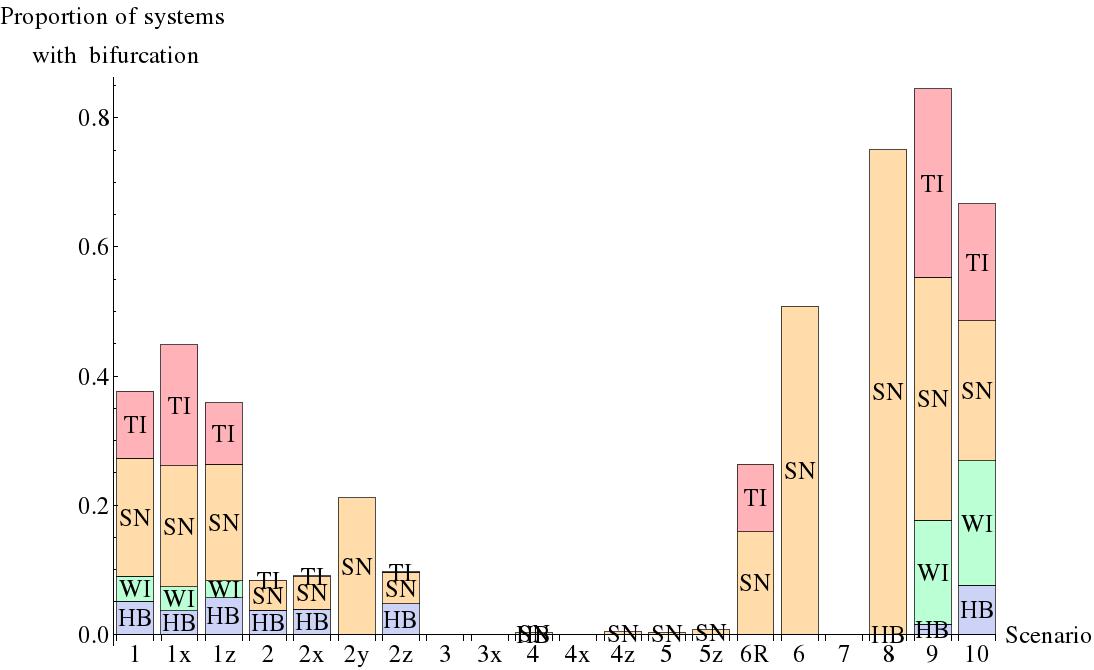} 
 \caption{The statistics of the four types of bifurcations for all considered models for $10^{5}$ random parameter sets. The numbers on the x-axis denote the model (see Tab.~\ref{tab:BigTabular}). Since more than one type of bifurcation can occur as a scale parameter is increased from 0 to 1,  the maximum total height of each bar is 4. The symbols are SN for saddle-node bifurcation, HB for Hopf bifurcation, TI for Turing instability and WI for wave instability.
(See Section \ref{sec:MSF} for the definition of the bifurcations.)}
 \label{fig:BifTotals}
\end{figure}

We evaluated how often each of the four different types of local bifurcations occurs as a migration scale parameter is increased from 0 to 1, averaging over  $10^{7}$  parameter sets and over the four different migration scale parameters. 
Varying the migration scale parameter means varying the relative contribution of migration to the population growth and loss terms. This approach is different from merely increasing a migration rate in an explicit model and therefore leads to somewhat different statistics of bifurcations. The comparison with explicit models will be provided in the next section.
The result is shown in  Fig.~\ref{fig:BifTotals}.

	The dominant bifurcation is always the saddle-node bifurcation. We see all four different types of bifurcations in models \emph{Stand} and  \emph{Standx} as well as models \emph{Growth} and \emph{Growth2}. These are the only models in which migration responds to population changes differently for predators and prey. The models with diffusive migration (\emph{Diff} and \emph{Jans}) do not show wave instabilities. For the scenarios of class II we see barely any bifurcations, with the exception of the models \emph{Huang},\emph{HuangR} and \emph{Growth2}, though model \emph{Huang} only exhibits saddle-node and model \emph{HuangR} saddle-node and Turing bifurcations. This is consistent with the earlier analysis of stability. If migration does not cause a change in stability, there can be no bifurcations that are induced by migration. 
Comparing these results with Table \ref{tab:AllCorr}, we see that there is some correlation between the amount by which migration decreases stability and the number of bifurcations that occur.

\FloatBarrier

\begin{figure}[htb]
\centering
 \includegraphics[width=0.7\textwidth]{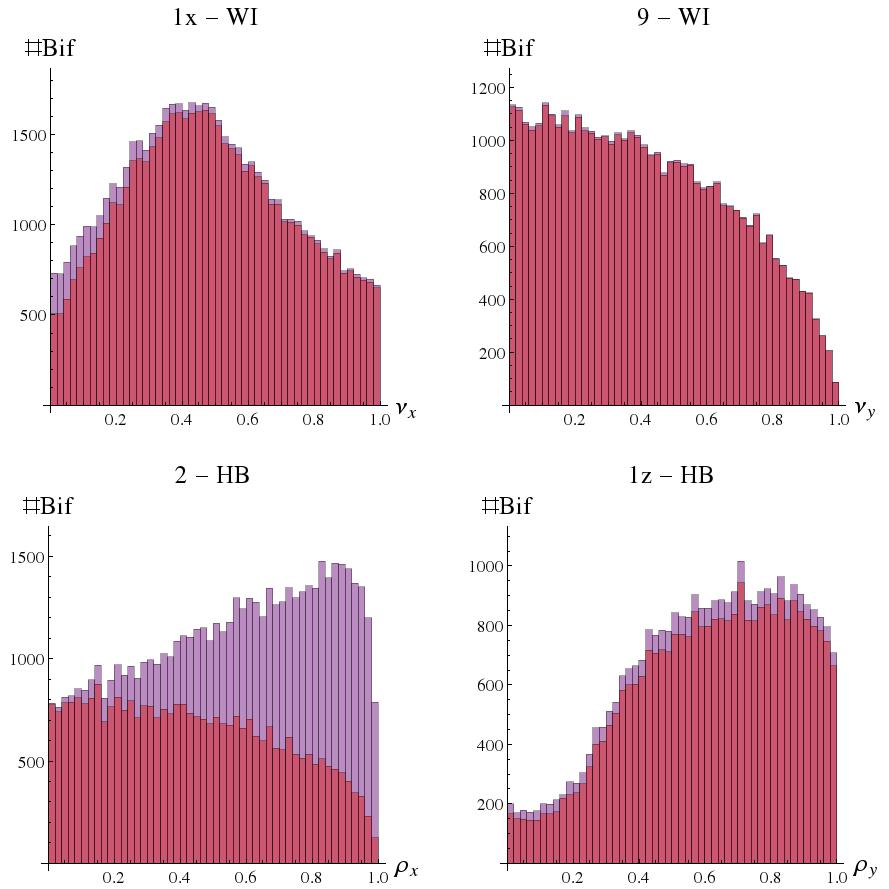} 
 \caption{The number of bifurcations into oscillating states in dependence of the migration scale parameters, in a sample of $10^{6}$ random parameter sets. The symbols above each graph denote the model (see Tab.~\ref{tab:BigTabular}) and type of bifurcation("HB" stands for synchronous Hopf bifurcation and "WI" stands for wave instability). The shape in the top left graph with most bifurcations at mid-range values is found for 38 percent of cases. Other forms that are seen less often (predominantly in models with complex migration dynamics) are shown in the other histograms.  The darker bars indicate the part of parameter sets with a bifurcation that would have been stable without any migration.}
 \label{fig:HopfSample}
\end{figure}

\begin{figure}[htb]
\centering
 \includegraphics[width=0.7\textwidth]{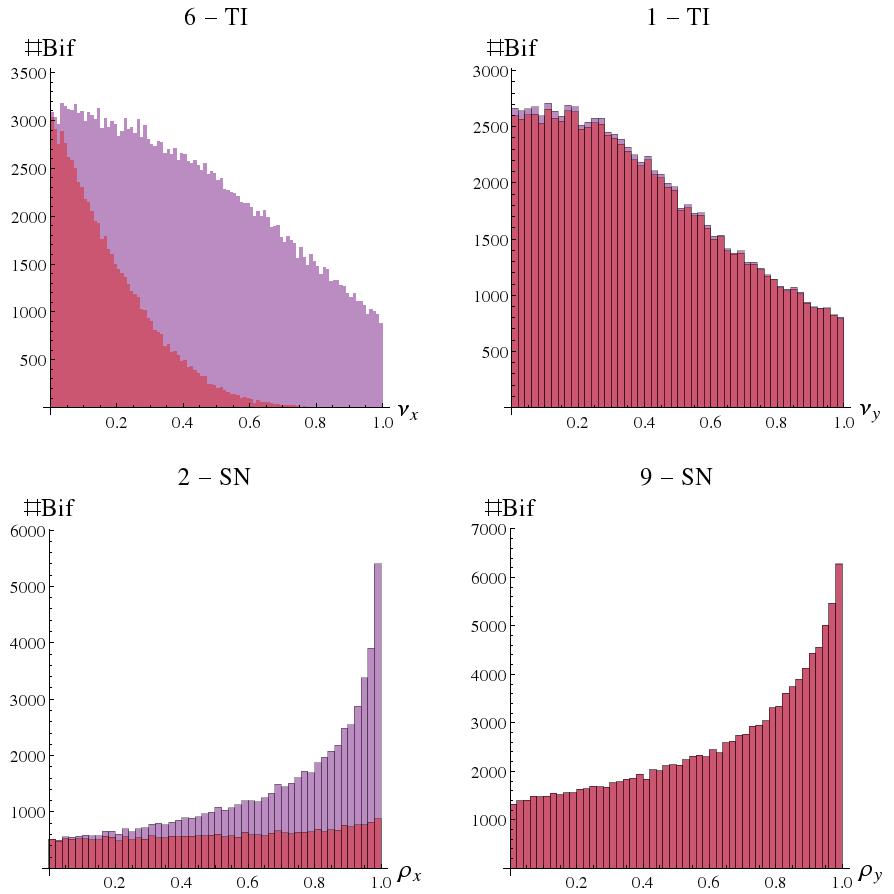} 
 \caption{Histograms of the number of saddle-node and Turing bifurcations over migrations scale parameters. The majority of bifurcations occur for scale parameters close to zero or 1. We sampled $10^{5}$ random sets of parameters per model.
 The number of bifurcations into oscillating states in dependence of the migration scale parameters, in a sample of $10^{6}$ random parameter sets. The symbols above each graph denote the model (see Tab.~\ref{tab:BigTabular}) and type of bifurcation("SN" stands for saddle-node bifurcation and "TI" stands for Turing instability). The majority of bifurcations, 67 percent of total cases, occur for scale parameters close to zero or 1. The darker bars indicate the part of parameter sets with a bifurcation that would have been stable without any migration.}
 \label{fig:SaddleSample}
\end{figure}

By evaluating the bifurcation statistics separately for different intervals of the migration scale parameters, we found that generally Hopf bifurcations and wave instabilities occur more often for intermediate values of the scale parameters, while saddle-node bifurcations and Turing instabilities occur more often for large and small values. Figs.~\ref{fig:HopfSample} and \ref{fig:SaddleSample} show the corresponding histograms for selected models. 
For these figures, we evaluated additionally the information  whether the used parameter set would produce a stable steady state without migration, i.e. with all migration scale parameters set to zero. This subset of the total number of bifurcations is marked by the violet colour.

Saddle-node bifurcations are more likely to occur at high migration values, whereas Turing-bifurcations tend to low and mid values. The steep increase of the number of saddle-node bifurcations for $\rho$ close to 1 is expected since large loss terms can lead to predator extinction, i.e., to a transcritical bifurcation, which is identical to a saddle-node bifurcation in the generalized modelling approach. Synchronous Hopf bifurcations are usually found at mid-to-high levels of migration while wave-instabilities prefer mid-to-low migration. Intermediate values of the migration scale parameters imply that the interactions between populations within and between patches are equally important for the dynamics, which in turn means that the phase space in which the dynamical trajectories of the system evolve is four-dimensional. Without migration or with very fast migration the phase space is only two-dimensional, with less complex dynamics and less oscillations.

It has been noted that prey pursuit or predator evasion can decrease synchrony \citep{li2005impact}. The scenarios \emph{ElAbdllaoui}, \emph{Growth} and \emph{Growth2} implement such types of migration. Model \emph{ElAbdllaoui} shows more synchronous Hopf bifurcations at high values of the migration scale parameters. Models \emph{Growth} and \emph{Growth2} have no clear trend. 

\FloatBarrier

\section{Comparison with results from explicit models}

 Many of the models listed in Table \ref{tab:BigTabular} were investigated earlier using explicit population dynamics models. These are the models by \citet{Mchich2007343},\citet{Jansen2001119,Jansen95}, \citet{Abdllaoui2007335}, \citet{Abrams201199} and \citet{huang2001predator}.  Our approach provides a different perspective and gives general insights. It thus  complement the findings of models that use specific functional forms for the different gain and loss terms. In order to illustrate the respective strengths of the different approaches, we compare in the following our results with those publications. We hereby have to keep in mind that our study was confined to homogeneous systems with identical patches and to local bifurcations of steady states, while the cited publications often include heterogeneous systems and global bifurcations as well.
 
 \citet{Jansen95} sees no dependence of the bifurcation condition on migration for the homogeneous steady state  with both species present on both patches. In our study, we found (few) saddle-node and Hopf bifurcations in this model when the migration scale parameters are changed. Such bifurcations are also present in the model by \citet{Jansen95} and are crossed when the parameters that characterize the in-patch dynamics are changed. This holds also for the follow-up paper from  2001\citep{Jansen2001119}. This nicely illustrates the fact that changing one parameter in the generalized approach is not equivalent to changing one parameter in a compatible explicit model, but to changing several parameters simultaneously, and vice versa. For instance, increasing the migration scale parameter $\nu$ in our approach means that a larger proportion of biomass increase is due to immigration, and that the biomass increase due to resource consumption decreases. In contrast, increasing the migration rate in an explicit model while keeping all other parameters fixed means that migration rate increases without a change in other processes. Varying scale parameters provides more generic insights as they refer to relative and not to absolute quantities. It is of course possible to change the parameters in an explicit model in a fixed relation and thus obtain the same kind of insight.
 
\citet{Mchich2007343} find a stabilizing effect of migration for a situation where the prey migrates with a rate that depends on predator density. This study uses a Lotka-Volterra model, for which a linear stability analysis is exact, and evaluates it in the limit of rapid migration. Our study always gives an eigenvalue zero of the Jacobian, since we look at homogeneous systems. The Hopf bifurcations observed by \citet{Mchich2007343} are due to the fact that they investigate the inhomogeneous case where the parameters on the two patches are different. The study by  \citet{Abdllaoui2007335} is similar to that by \citet{Mchich2007343}, but uses a type II functional response and finds that migration can create limit cycles. We see a large percentage of stable webs, with Hopf bifurcations occurring only rarely. This leads to the conclusion that the majority of Hopf bifurcations seen by  \citet{Abdllaoui2007335} is due to the fact that they study an inhomogeneous system. 

\citet{Abrams201199} state that adaptive migration produces frequent anti-synchronous limit cycles. Our study generally shows wave instabilities for scenarios with adaptive migration (models \emph{Stand}, \emph{Growth}), and such wave instabilities lead to anti-synchronous limit cycles.  However, we did not find wave instabilities for model \emph{Abrams}. Since \citet{Abrams201199} do not show bifurcation diagrams, there is no contradiction between their and our findings. They focussed on parameter ranges where local patch dynamics is oscillatory. It is well possible that in this model antisynchronous oscillations result only (or almost always) out of inhomogeneous fixed points or by non-local bifurcations. 

\citet{huang2001predator} investigate the case that predators cannot migrate while handling prey, which means that predators migrate at low prey densities but become immobile for large prey densities. By performing a linear stability analysis, the authors found a stable steady state  for a wide range of parameters, which can become unstable by Hopf bifurcations and saddle-node bifurcations. We only find saddle-node bifurcations when we vary the migration scale parameters. However, we find also Hopf bifurcations when we vary additionally one of the local parameters. This again illustrates the fact that changing one parameter in an explicit model corresponds to changing several parameters in the generalized model, and vice versa. Furthermore, \citet{huang2001predator} state that antisynchronous oscillations are always unstable if they exist, in agreement with our result that a stable steady state cannot become unstable by a wave instability. 

To conclude this section, we would like to point out that a model with explicit population dynamics explores only a small subspace of the general class of models that are compatible with what is known from empirical systems. In particular, it introduces functional dependencies between the parameters used in the generalized modelling approach and thus limits the space that can be explored. Other explicit models would lead to somewhat different functional relations between the coresponding generalized parameters and would thus explore a different region of the space of possibilities. For instance, the model by \citet{Abrams201199} does not show wave instabilities, which do however occur in models that belong to the same overall class of systems that have a growth-rate dependent migration. Also, the model by \citet{Mchich2007343} finds Hopf bifurcations only in inhomogeneous systems, due to the Lotka-Volterra form of local population dynamics, while a generalized investigation that does not insist on this unrealistic constraint shows that migration can generically induce Hopf bifurcations in homogeneous systems. Of course, we cannot rule out on the other hand that the generalized approach includes parameter combinations that cannot be satisfied by any realistic and explicit population dynamics model.

\section{Conclusions}

In this paper we studied the impact of dispersal on a general class of 2-patch 2-species predator-prey system. 
We used the approach of generalized modelling, which allows to investigate the
stability of the system without restricting the kinetics to specific functional forms.
In comparison to previous studies  we are thus able to provide a broader overarching perspective.  
  
Our analysis confirms that dispersal generally decreases the local stability of the system. 
Although, dispersal creates benefits, such as the rescue effect, not studied here, it does not generally promote the dynamical stability of the system. Thus dispersal may lead to undesirable dynamics, causing for instance increased spatial, temporal, or spatio-temporal variability. However, we found also large parameter regions in which dispersal increases the stability and may thus help avoiding such undesirable dynamics.  
 
For the case of identical patches we were able to compute the thresholds at which destabilization occurs analytically. Furthermore, we used a numerical sampling procedure
for a broad survey of the impact of parameters of the generalized model. By restricting the parameters of the generalized model to appropriate ranges we were able to analyse 19 different scenarios. 

In a number of scenarios the impact of dispersal is very weak. For instance if dispersal occurs completely randomly it does not have a notable effect on stability. Moreover, density independent dispersal only affects stability if the growth of the prey shows effects of saturation. Finally, superlinear mortality rates, such as quadratic mortality, have a stabilizing effect that can be much stronger than the effect of dispersal, thus that dispersal does not generally lead to a destabilization of such systems. These findings are consistent with \citet{TromeurRudolfGross2013}, who studied the dynamics of a single population in a system with many patches. In contrast to this previous study we found that dispersal can in certain cases be stabilizing in the two-species system. The parameter regions in which a positive effect of dispersal on stability are observed are much wider when dispersal of a given species depends on the densities of other species in the system.  

While we studied only 2 patches, our results allow for some extrapolation to larger systems. 
We need to distinguish between pattern forming (Turing, wave) and non-pattern forming (Hopf, saddle-node) instabilities. In agreement with \citet{Abrams201199} we find that pattern forming instabilities can only occur in systems with non-random dispersal, a result that should hold also in systems with more patches. 

Using general insights into the dynamics of networks \citep{do2012engineering,macarthur2008symmetry} we can say that non-pattern-forming instabilities cannot depend on network structure.
If dispersal affects these instabilities then it does so only because it shifts the operating point of the system and introduces new nonlinearities, e.g.~from losses during migration. Both of these effects can be captured faithfully in single patch models, where migration affects are modelled as additional terms.  

By contrast, pattern forming instabilities depend sensitively on network structure and thus can only be analysed if the spatial structure at hand is captured in the model. Previous results\citep{do2012engineering} suggest that the symmetric pair of patches studied here is a particularly unstable configuration that promotes pattern-forming instabilities, while larger, less symmetric systems should be tentatively more stable. 

Perhaps most importantly our findings illustrate the complex nature of dispersal effects.
In the simplest scenarios dispersal is neither stabilizing nor destabilizing, but the interplay of dispersal with other nonlinearities in the system can increase or reduce stability. To fully understand dispersal effects in complex food webs, and in particular pattern forming instabilities, we will eventually have to study large, many-patch, many-species systems. For addressing this challenge, the generalized modelling approach, used here, could be valuable tool.

\section*{Acknowledgements}
We thank Eric Tromeur for helping with the initial literature search and Andreas Brechtel for useful discussions and cross-checking computational data. This work was supported by DFG under contract number Dr300/12-1.

\bibliography{science}
 
\bibliographystyle{elsarticle-harv}

\end{document}